\documentclass[preprint,10pt,number,5p,twocolumn]{elsarticle}
\usepackage[utf8]{inputenc}
\usepackage{float}
\usepackage{graphics}

\usepackage[footnotesize]{caption2} 

\usepackage{threeparttable} 
\usepackage{multirow}
\usepackage{flushend} 
\usepackage{verbatim}
\usepackage{graphicx}
\usepackage{color}
\usepackage{amsmath,amssymb}
\usepackage{amsmath}
\usepackage{hyperref}
\DeclareUnicodeCharacter{00A0}{~}

\begin{document}
\title{Isospin-dependence of the charge-changing cross-section shaped by the charged-particle evaporation process}

\author[a,b,c]{J.W. Zhao}
\author[a]{B.-H. Sun\corref{cor}}
\ead{bhsun@buaa.edu.cn}
\author[a,d]{I. Tanihata}
\author[a]{S. Terashima}
\author[ap]{A. Prochazka}
\author[a]{J.Y. Xu}
\author[a]{L.H. Zhu}
\author[b]{J. Meng}
\author[js]{J. Su}
\author[b,kyz]{K.Y. Zhang}
\author[a]{L.S. Geng}
\author[a]{L.C. He}
\author[a]{C.Y. Liu}
\author[a]{G.S. Li}
\author[els]{C.G. Lu}
\author[a]{W.J. Lin}
\author[wpl]{W.P. Lin}
\author[els,zl]{Z. Liu}
\author[wpl]{P.P Ren}
\author[els]{Z.Y. Sun}
\author[a]{F. Wang}
\author[a]{J. Wang}
\author[a]{M. Wang}
\author[els]{S.T. Wang}
\author[a]{X.L. Wei}
\author[els]{X.D. Xu}
\author[a]{J.C. Zhang}
\author[a]{M.X Zhang}
\author[els]{X.H. Zhang}

\cortext[cor]{Corresponding author.} 
\address[a]{School of Physics, Beihang University, Beijing, 100191, China}
\address[b]{State Key Laboratory of Nuclear Physics and Technology, School of Physics, Peking University, Beijing, 100871, China}
\address[c]{GSI Helmholtzcentre for Heavy Ion Research GmbH, Darmstadt, 64291, Germany}
\address[d]{RCNP, Osaka University, Ibaraki Osaka 567-0047, Japan
}
\address[ap]{Medaustron, 2700 Wiener Neustadt, Austria}

\address[js]{Sino-French Institute of Nuclear Engineering and Technology, Sun Yat-sen University, Zhuhai, 519082, China}
\address[kyz]{Institute of Nuclear Physics and Chemistry, CAEP, Mianyang, Sichuan, 621900, China}

\address[els]{Institute of Modern Physics, Chinese Academy of Sciences, Lanzhou, 730000, China}

\address[wpl]{Key Laboratory of Radiation Physics and Technology of the Ministry of Education, Institute of Nuclear Science and Technology, Sichuan University, Chengdu, 610064, China}
\address[zl]{School of Nuclear Science and Technology, University of Chinese Academy of Sciences, Beijing, 100049, China}

\date{June 2023}

\begin{abstract}
	We present the charge-changing cross sections (CCCS) of $^{11-15}$C, $^{13-17}$N, and $^{15,17-18}$O at around 300 MeV/nucleon on a carbon target, which extends to $p$-shell isotopes with $N < Z$ for the first time. The Glauber model, which considers only the proton distribution of projectile nuclei, underestimates the cross sections by more than 10\%. We show that this discrepancy can be resolved by considering the contribution from the charged-particle evaporation process (CPEP) following projectile neutron removal. Using nucleon densities from the deformed relativistic Hartree-Bogoliubov theory in continuum, we investigate the isospin-dependent CPEP contribution to the CCCS for a wide range of neutron-to-proton separation energy asymmetry. Our calculations, which include the CPEP contribution, agree well with existing systematic data and reveal an ``evaporation peak" at the isospin symmetric region where the neutron-to-proton separation energy is close to zero. These results suggest that analysis beyond the Glauber model is crucial for accurately determining nuclear charge radii from CCCSs. 
\end{abstract}

\begin{frontmatter}
\end{frontmatter}

\section{Introduction}
\label{intro}
Nuclear reactions at various energies have provided valuable insights into quantum chromodynamics, atomic nuclear structure, the origin of elements, and the properties of neutron stars. The use of radioactive ion beams created by nuclear reactions has greatly expanded our knowledge from stable, sphere-like nuclei to the exotic shapes and phenomena exhibited by nuclei~\cite{Tanihata85}. At intermediate energies above 200 MeV/nucleon, nuclear reactions can be well described by nucleon-nucleon collisions~\cite{Glauber59}. Nuclear interaction and charge-changing cross sections are the most straightforward observables in nuclear reactions. They allow us to deduce the size of nuclei, visualize their structure (including skin/halo and shell closure~\cite{Blank91, tanihat2013}), 
test/challenge nuclear theories, and understand the equation of state of asymmetric nuclear matter~\cite{Aumann17, Xu22}. 

The charge-changing cross section (CCCS) describes the probability of an incident nucleus losing protons by interacting with the target. 
The data have been found to be well correlated with  
the root-mean-square proton distribution radii ($R_p$) of stable and exotic nuclei in combination with the Glauber model, which considers only the collision probability of projectile protons with the target nucleus, while projectile neutrons are treated as spectators~\cite{Meng02, AB04}. However, precision measurements show that the CCCS does not always provide a direct probe of the proton radius, and essential corrections~\cite{Yama10, Yama11, Ritu16} are often required to extract $R_p$ of unstable nuclei from their experimental CCCSs. For instance, a phenomenological scaling factor must be introduced to the zero-range optical-limit Glauber model (ZRGM) to determine the proton radii of $^{15,16}$C~\cite{Yama10, Yama11}, $^{30}$Ne, and $^{32,33}$Na~\cite{Ozawa13} from CCCS measurements at around 300 MeV/nucleon. In contrast, deducing the proton radii of isotopes such as Be~\cite{Tera14}, B~\cite{GSI14B}, C~\cite{Ritu16, Tran16}, N~\cite{GSI19N}, O~\cite{Kaur22}, and Ca~\cite{Tanaka21} from CCCS measurements at different energy regions requires fine adjustments to the Glauber model. This results in inconsistent treatments that limit our understanding of the reaction mechanism. One possible approach to address this ambiguity is to consider the contribution of the charged-particle evaporation from the intermediate pre-fragments in addition to the Glauber model~\cite{Tanaka21}. However, the extent to which this evaporation process contributes to the CCCS and its dependence on the reaction energy, projectile masses, and proton-neutron ratios remain unclear.  

This Letter presents the results of the first charge-changing cross section (CCCS) measurement campaign at the Heavy Ion Research Facility in Lanzhou (HIRFL-CSR), China. Thirteen CCCSs were determined for stable and unstable nuclei ($^{11-15}$C, $^{13-17}$N, and $^{15,17-18}$O) on carbon using the transmission method at 300 MeV/nucleon. The reaction is highly sensitive to the nucleon density distributions of interest due to the significant difference in proton-proton and proton-neutron cross sections at this energy region. These measurements represent the first extension of CCCSs for C, N, and O isotopes to $N<Z$ nuclei. This data completes the CCCSs of $p$-shell nuclei near the $\beta$-stability line at 300 MeV/nucleon and covers a wide range of neutron-proton separation energies up to 40 MeV, which is crucial for a better understanding of the underlying reaction mechanism. 

\section{Experiment and analysis}
\label{expana}
The experiment was performed at the External Target Facility (ETF), located alongside the third focus (F3) of the second Radioactive Ion Beam Line (RIBLL2) at HIRFL-CSR~\cite{Xia02}. A comprehensive description of RIBLL2 and ion-optics of RIBLL2-ETF can be found in Ref.~\cite{Sun18}.
Secondary beams were produced by the fragmentation of $^{18}$O projectiles at 400 MeV/nucleon interacting with a 
5.54 g/cm$^2$ thick Be target. Fragments were separated in flight by the first half (F0-F2) of RIBLL2 and then delivered to ETF. Nuclei of interest were identified event-by-event by using the magnetic rigidity ($B\rho$), time-of-flight (TOF), and energy loss ($\Delta E$) information. A 2.77 g/cm$^2$ thick carbon plate was used as the reaction target at ETF. The TOF resolution is better than 100 ps ($\sigma$)~\cite{Lin17}. 
A multiple-sampling ionization chamber (MUSIC) and a silicon stack were used to measure $\Delta E$ of the incident and reaction particles, respectively. The particles were tracked using two multi-wire proportional chambers before the reaction target and a third one after it. The charge and position resolutions are better than 0.25 ($\sigma$)~\cite{Zhang15, Zhao19} and 0.5 mm ($\sigma$), respectively. 

In the transmission measurement, we counted the numbers of the incident and reaction particles without losing protons (i.e., $Z$-unchanged particles). To account for reactions in materials other than the target, such as detectors, we performed a measurement without the target using the same setup. The CCCS is then calculated using the equation $\sigma_{\rm cc}=-(1/t)\ln(\gamma/\gamma_{0})$, where $\gamma$ and $\gamma_{0}$ are the ratios of $Z$-unchanged particles for the cases with and without the target, respectively. $t$ is the number of target nuclei per unit area. The contamination level of incident particles was always less than $10^{-4}$. The statistic errors of CCCSs and the systematic error are typically 3\% and 0.5\%, respectively. The systematic error is estimated by examining the energy loss spectra of the silicon stack as well as their correlation. 
The statistical error and systematic error are added in quadrature to obtain the total error. For the details in data analysis, we refer to Ref.~\cite{Wang23}. 

\section{Results and discussions}
\begin{figure}[htb]
	\centering
	\includegraphics[width=0.4\textwidth]{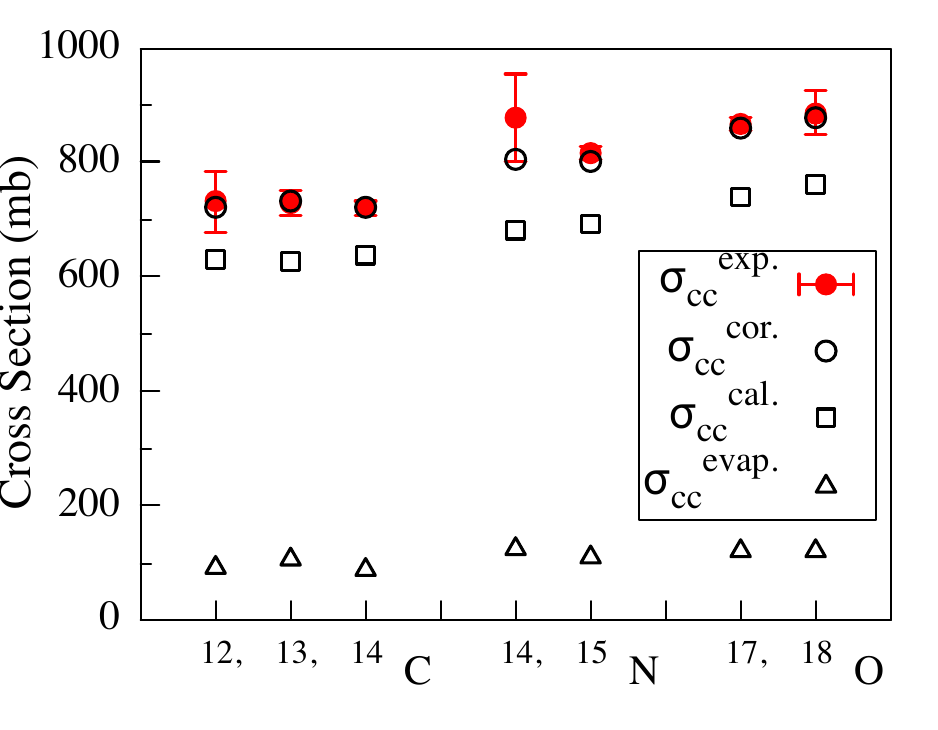}
	\caption{CCCSs of $^{12-14}$C, $^{14-15}$N and $^{17-18}$O on carbon (filled circles) measured in this work. The open squares represent the theoretical predictions with the ZRGM. Considering the contribution from the charged-particle evaporation (open triangles), the corrected values (open circles) can well reproduce the data.}
	\label{f1_exp_data}
\end{figure}

As summarized in Table~\ref{tab1}, our results are in good agreement with those obtained at similar energies~\cite{Yama11}. 
The ZRGM predicted CCCSs for $^{12-14}$C, $^{14-15}$N and $^{17-18}$O are shown for comparison. Nuclear charge radii of these nuclei are well-known from the electron scattering ~\cite{Angeli13}. In the ZRGM calculation, we employed the harmonic oscillator type of proton densities to reproduce the charge radii of these nuclei ~\cite{Angeli13}, while the projectile neutrons are treated as spectators. As shown in Fig.~\ref{f1_exp_data}, the ZRGM predictions ($\sigma _{\rm cc}^{\rm cal.}$) systematically underestimate the experimental data ($\sigma _{\rm cc}^{\rm exp.}$) by 11-15\%. 

\begin{table}[htb]
\centering
\vspace{-0.2cm}
\caption{CCCSs measured in this work. The available data (Lit.) at around 300 MeV$/$nucleon from Ref.~\cite{Yama11} and ZRGM predictions are shown for comparison.}
\label{tab1}
\begin{tabular}{ p{0.5cm} p{2.2cm} p{1.3cm} p{1.3cm} p{1.3cm} }
\hline
\hline
 & Energy (MeV/nucleon) & CCCS (mb) & Lit. (mb) & ZRGM (mb)\\
\hline
 $^{11}$C & 319 (38) & 716 (20) &     &     \\    
 $^{12}$C & 294 (5)  & 731 (52) &     &    631    \\ 
 $^{13}$C & 322 (30) & 729 (22) &     &    628     \\
 $^{14}$C & 339 (12) & 732 (22) & 731 (5) & 637 \\
 $^{15}$C & 327 (16) & 758 (56) & 743 (6) &  \\
 $^{13}$N & 310 (29) & 752 (35) &     &         \\
 $^{14}$N & 289 (4)  & 878 (77) &     &   683      \\
 $^{15}$N & 315 (21) & 815 (11) &     &   691  \\
 $^{16}$N & 322 (18) & 813 (9)  &     &         \\
 $^{17}$N & 328 (13) & 790 (11) &     &         \\
 $^{15}$O & 301 (24) & 880 (18) &     &         \\
 $^{17}$O & 308 (12) & 866 (11) & 896 (9) & 741 \\
 $^{18}$O & 368 (2)  & 887 (39) & 891 (10) & 760\\
\hline
\end{tabular}
\end{table}

When neutrons are removed from the projectile in collisions at relativistic energies, the residual nucleus can be excited to unbound states higher than the charged-particle emission threshold. 
The de-excitation by emitting charged particles (e.g., proton, $\alpha$) will contribute to the charge-changing reaction. This charged-particle evaporation process (CPEP) is not accounted for in the present Glauber model calculation and should be considered to interpret the experimental data~\cite{ Tanaka21}. Extreme examples are the proton drip line nuclei like $^{17}$Ne, removing neutron(s) from it will lead to unbound pre-fragments, which decay by emitting proton(s). In these cases, the CCCS is the same as its reaction cross sections due to the CPEP~\cite{Moriguchi20}.

The CCCS rooted in CPEP can be calculated as

\begin{equation}
\sigma_{\rm cc}^{\rm evap.}=\sum_{i} \sigma_{-i\rm n} \cdot P_{i}^{\rm evap.}\;,
\label{eq: CPEP}
\end{equation}
where $\sigma_{-i\rm n}$ is the cross section of only removing $i$ neutrons. One needs to note that this $\sigma_{-i\rm n}$ and the experimental neutron removal cross sections ($\sigma_{-i\rm {n}}^{\rm {exp.}}$) are not directly comparable. The $\sigma_{-i\rm {n}}^{\rm {exp.}}$ containing only the population of the ground and bound states that are experimentally accessible,
but the $\sigma_{-i\rm n}$ here contains additionally the population of unbound states. 

$P_{i}^{\rm evap.}$ refers to the corresponding charged-particle evaporation probability, which is strongly dependent on the excitation energy distribution (EED) of the corresponding neutron-removed pre-fragment. The parameter-adjusted Gaimard-Schmidt (GS) approach~\cite{GS91, Schei09} is often used to estimate EEDs. In addition, the isospin-dependent quantum molecular dynamics model (IQMD)~\cite{Su11} provides a microscopic-rooted EED, which takes into account the reaction dynamics and incident energy dependence. Calculations of $\sigma_{-i\rm n}$ and $P_{i}^{\rm evap.}$ are further detailed in the Supplemental Material. 

Figure~\ref{f1_exp_data} presents the $\sigma _{\rm cc}^{\rm evap.}$ for $^{12-14}$C, $^{14-15}$N and $^{17-18}$O calculated with the EEDs from the GS approach.  
The $\sigma _{\rm cc}^{\rm cor.}$, i.e., the sum of $\sigma _{\rm cc}^{\rm evap.}$ and the ZRGM predictions ($\sigma _{\rm cc}^{\rm cal.}$, same as listed in Table~\ref{tab1}), can well reproduce our data. For stable $p$-shell isotopes, the CPEP contributions are almost constant and account for more than 10\% of the experimental CCCSs. 

To see the global behavior, we calculated $\sigma _{\rm cc}^{\rm evap.}$ in a wide isospin range at 300 MeV/nucleon for C, N, O, Si, Ar, and Ca isotopes, of which systematic experimental data are available over a large variation of neutron-to-proton separation energy asymmetry. The nucleon densities are from the deformed relativistic Hartree-Bogoliubov theory in continuum (DRHBc)~\cite{Zhou10,DRHBc, Zhang20} with the density functional PC-PK1~\cite{Zhao10}. This state-of-art model predicts reasonable ground-state properties of even-even nuclei with 8 $\le Z \le$ 120 from the proton drip line to the neutron drip line~\cite{Zhang22}. In addition to EEDs from the GS approach, we performed calculations for Ca isotopes using EEDs from the IQMD model as well. 

\begin{figure}[t]
	\centering
       \includegraphics[width=0.5\textwidth]{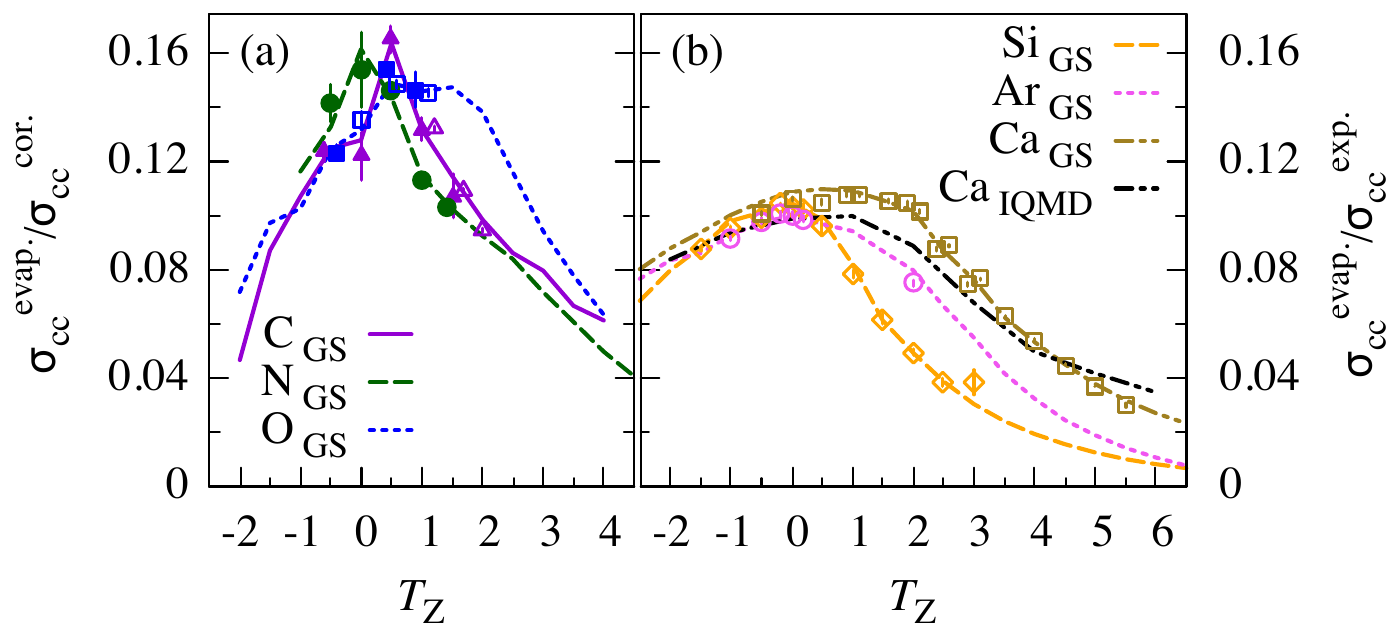}
	\caption{(a) CPEP contribution ratios to CCCSs ($\sigma _{\rm cc}^{\rm cor.}$, lines) and to experimental results of C (filled triangles from this work and open triangles taken from~\cite{Yama11}), N (filled circles from this work), and O (filled squares from this work and open squares taken from~\cite{Yama11}) isotopes on $^{12}$C at 300 MeV/nucleon. EEDs are from the GS approach. (b) Same as (a) but for Si (open diamonds, data taken from~\cite{Sawahata17,LiSun}), Ar (open circles, data taken from ~\cite{Sawahata17,Iancu05}), and Ca (open square, data taken from~\cite{Tanaka21, Yamaki13}) isotopes. EEDs are from the GS approach and IQMD model.}
	\label{ratio_isospin}
 \vspace{-0.2cm}
\end{figure}

As shown in Fig.~\ref{ratio_isospin} (a), the CPEP contribution ratios to CCCSs of the $p$-shell nuclei, $\sigma _{\rm cc}^{\rm evap.}/\sigma _{\rm cc}^{\rm cor.}$, exhibit a clear isospin dependence with a maximum at around $T_Z=0$. This isospin dependence is confirmed by the CPEP contribution ratios to our data, $\sigma _{\rm cc}^{\rm evap.}/\sigma _{\rm cc}^{\rm exp.}$, as shown with symbols in Fig.~\ref{ratio_isospin} (a). The CPEP is suppressed towards the isospin asymmetric side and tends to be negligible for the very neutron-rich isotopes at $T_Z>5$. Similar isospin dependence behavior is also observed for Si, Ar, and Ca isotopes in terms of the $\sigma _{\rm cc}^{\rm evap.}/\sigma _{\rm cc}^{\rm cor.}$ as shown in Fig.~\ref{ratio_isospin} (b). The maxima for Si, Ar, and Ca isotopes are about 20\% smaller than those for $p$-shell isotopes. This suggests a weaker CPEP contribution towards the heavier isotopes. Moreover, the ``width" of the distribution of $sd$-shell isotopes is broader than that for $p$-shell isotopes. This is mainly due to the more pronounced single-particle effect in the $p$-shell nuclei, which is also reflected by the less smooth isospin dependence of the CPEP contribution ratio.

In Fig.~\ref{ratio_isospin}, we compare $\sigma _{\rm cc}^{\rm evap.}/\sigma _{\rm cc}^{\rm cor.}$ (shown with lines) with $\sigma _{\rm cc}^{\rm evap.}/\sigma _{\rm cc}^{\rm exp.}$ (shown with symbols) for C, N, O, Si, Ar, and Ca isotopes. Their agreement indicates that the corrected results ($\sigma _{\rm cc}^{\rm cor.}$) can reproduce the experimental CCCSs measured in this work and data taken in Refs.~\cite{Yama11, Sawahata17, Tanaka21, Yamaki13, Iancu05, LiSun} within the experimental uncertainty. In particular, the calculations employing EEDs from IQMD show very good prediction power for Ca isotopes. It is worth noting that the projectile excitation in IQMD is driven by the microscopic interaction between nucleons. No parameters were specially adjusted for EEDs.

\begin{figure}[tbh]
	\centering
	\includegraphics[width=0.5\textwidth]{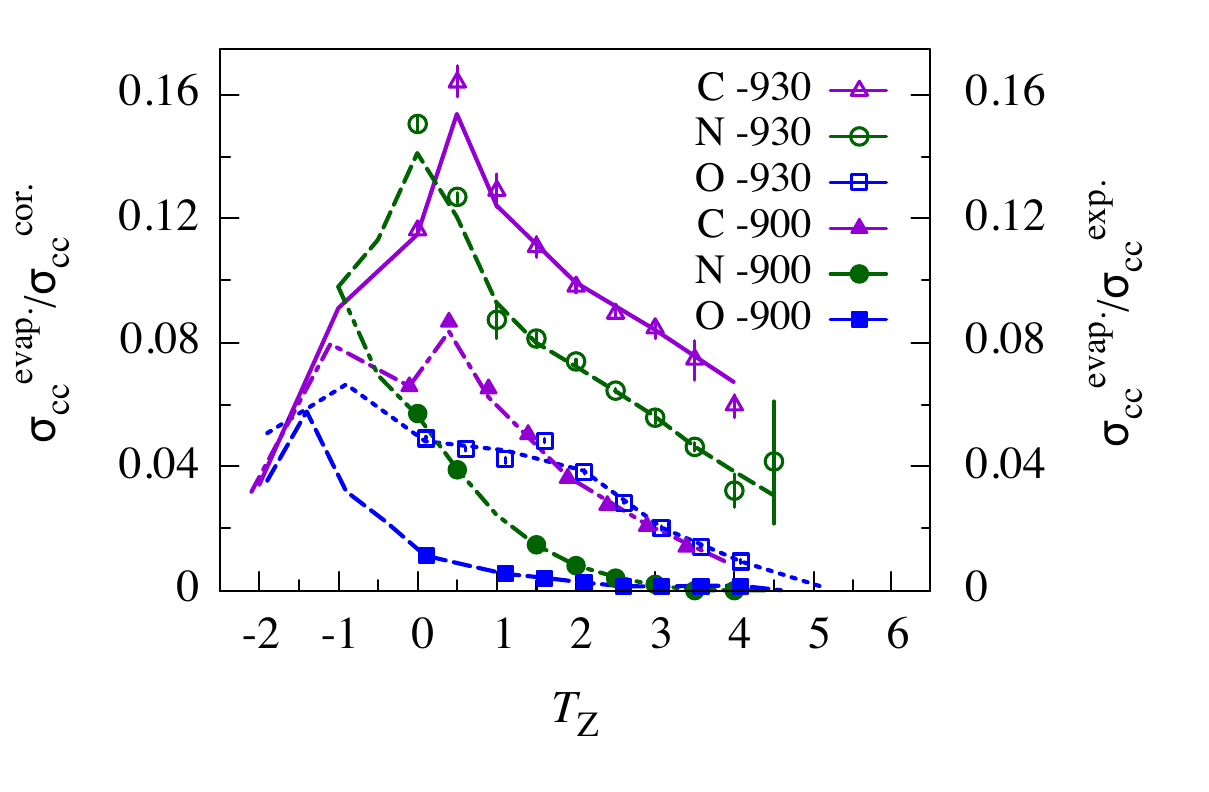}
	\caption{CPEP contribution ratios to CCCSs ($\sigma _{\rm cc}^{\rm cor.}$) of C, N and O isotopes on $^{12}$C at around 900 MeV/nucleon. Estimations with EEDs from the GS approach are shown with lines. CPEP contribution ratios to experimental data measured at 900 MeV/nucleon~\cite{Ritu16, GSI19N, Kaur22} and 930 MeV/nucleon~\cite{Chulkov00} are shown with symbols.}
	\label{evap_isospin_900}
\end{figure}
 
We extended our calculations to compare with the experimental CCCSs of C, N, and O isotopes measured at 900 MeV/nucleon~\cite{Ritu16, GSI19N, Kaur22} and 930 MeV/nucleon~\cite{Chulkov00}. The two sets of data have systematic deviations up to 40 mb and will thus affect the corresponding parameter optimization for EEDs from the GS approach. 
Nevertheless, a similar isospin dependence pattern is seen on the $\sigma _{\rm cc}^{\rm evap.}/\sigma _{\rm cc}^{\rm exp.}$ as 
shown in Fig.~\ref{evap_isospin_900}. This suggests that the isospin-dependent behavior is a common feature for reactions at 300 and 900 MeV/nucleon.

\begin{figure}[t]
	\centering
    \includegraphics[width=0.5\textwidth]{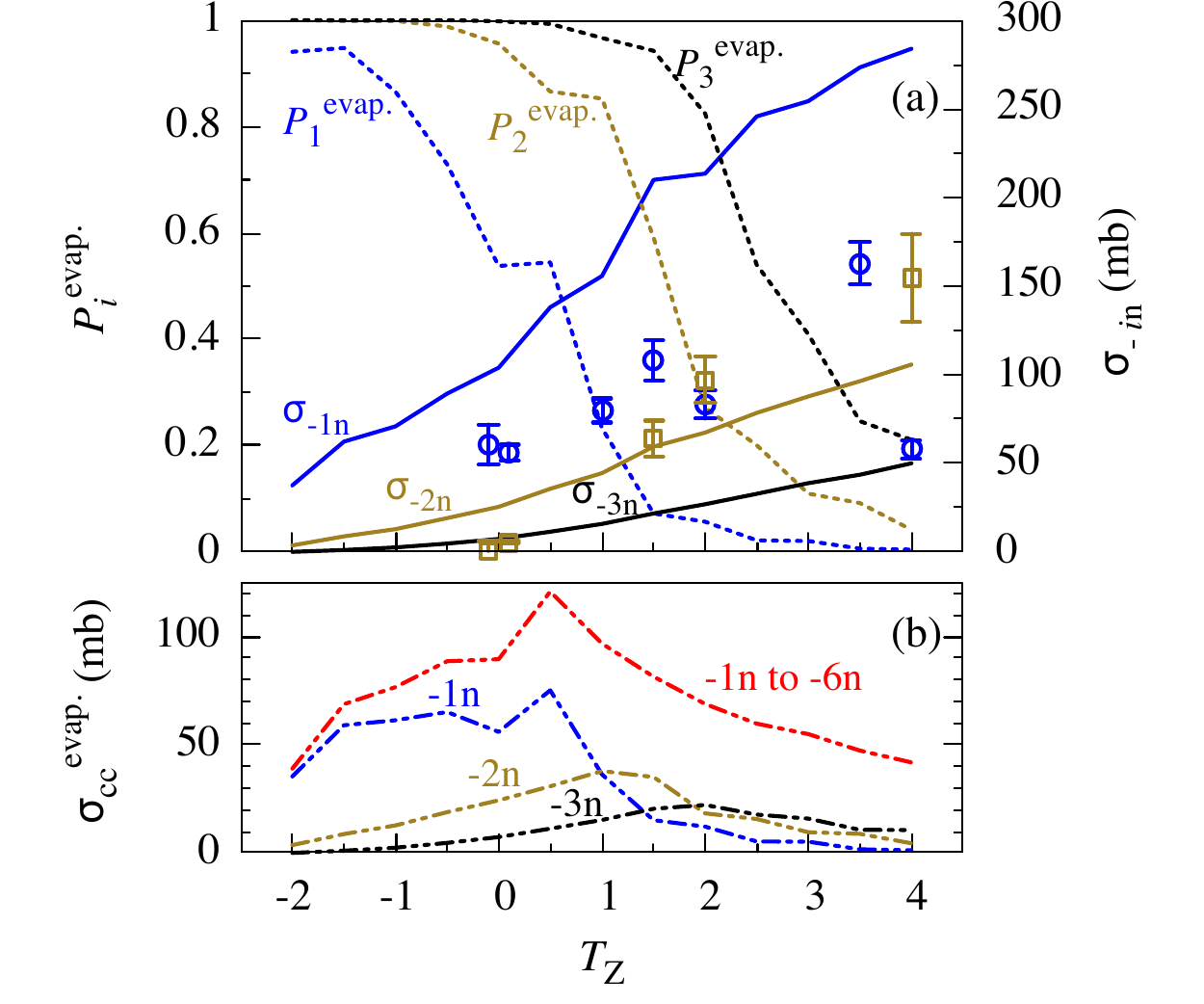}
	\caption{(a) One to three neutrons removal cross sections ($\sigma_\text{-in}$) of $^{10-20}$C interacting with $^{12}$C at 300 MeV/nucleon and the corresponding charged-particle evaporation probabilities ($P_\text{i}^\text{evap.}$ ) of the neutron removed pre-fragments. Experimental one- and two-neutron removal cross sections measured at 240 MeV/nucleon are shown for comparison by open circles and open squares~\cite{Sun20,Kidd88,Sun21,Koba12}, respectively. (b) The CPEP contribution from one to three neutrons removal and the total CPEP contribution considered up to six neutrons removal.}
	\label{C_evap_xn_isospin}
\end{figure}

To understand the nature of such isospin dependence, we calculated the cross sections of removing only neutrons (i.e., $\sigma_{-i\text{n}}$) from $^{10-20}$C 
when colliding with $^{12}$C at 300 MeV/nucleon and the corresponding charged-particle evaporation probabilities (i.e., $P_{i}^\text{evap.}$) from the pre-fragments right after neutron removals. The cases of removing up to three neutrons are shown in Fig.~\ref{C_evap_xn_isospin} (a) as they contribute more than 90\% to the final $\sigma_{\rm cc}^{\rm evap.}$ at $T_z \sim 0$. As seen in Fig.~\ref{C_evap_xn_isospin} (a), despite the bump at $^{13}$C due to the single-particle effect, $\sigma_{-i\text{n}}$ increases with isospin due to the increasing neutron-proton ratio and the decreasing trend of neutron separation energies. 
$P_{i}^\text{evap.}$ is strongly suppressed towards the very neutron-rich side due to the increasing of charged-particle separation energies~\cite{AME20}. 
Consequently, $\sigma_{\rm cc}^{\rm evap.}$ (see Eq.~\ref{eq: CPEP}) considering up to six neutron removal as a whole decrease at larger isospin as shown in Fig.~\ref{C_evap_xn_isospin} (b). As for the proton-rich projectile, $\sigma_{\rm cc}^{\rm evap.}$ is dominated by the small $\sigma_{-i\text{n}}$ and is predicted to be suppressed as well. 

Experimental one- and two-neutron removal cross sections~\cite{Sun20,Kidd88,Sun21,Koba12}, $\sigma_{\rm {-1n}}^{\rm {exp.}}$ and $\sigma_{\rm {-2n}}^{\rm {exp.}}$ of $^{12,14,15,16,19,20}$C measured at 240 MeV/nucleon are shown in Fig.~\ref{C_evap_xn_isospin} (a) for comparison. The $\sigma_{\rm {-1n}}^{\rm {exp.}}$ and $\sigma_{\rm {-2n}}^{\rm {exp.}}$ are inclusive cross sections measured by knockout reactions that contain only the population of the ground and bound states. In contrast, the $\sigma_{-i\rm n}$ contains additionally the population of unbound states. Therefore, $\sigma_{-1\rm n}$ values are twice larger than the corresponding $\sigma_{-1\rm n}^{\rm {exp.}}$, while the $\sigma_{-2\rm n}^{\rm {exp.}}$ values are underestimated by about 30\%. The extreme case due to the population of unbound states is seen for $^{20}$C. As discussed in Ref.~\cite{Koba12}, the population of unbound states in $^{19}$C by removing one neutron from $^{20}$C and the decay of these unbound states by one neutron emission makes the $\sigma_{\rm {-2n}}^{\rm {exp.}}$ unusually larger than its $\sigma_{\rm {-1n}}^{\rm {exp.}}$.

\begin{figure}[t]
	\centering
    \includegraphics[width=0.5\textwidth]{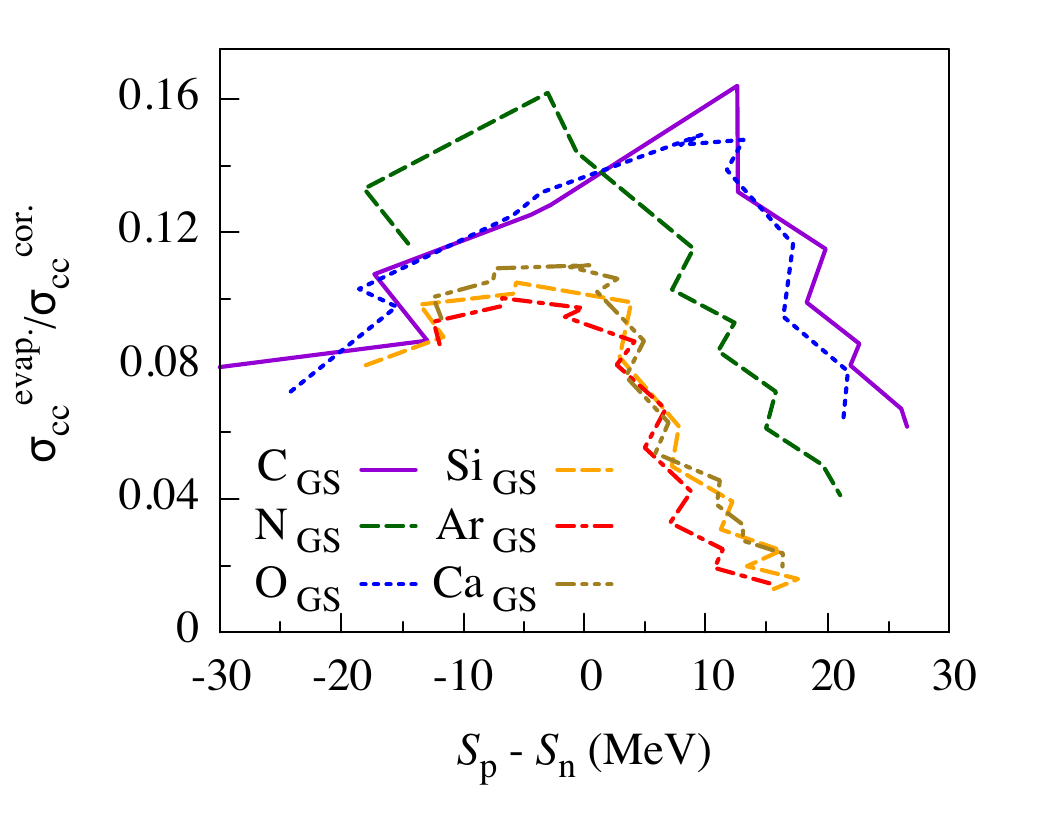}
	\caption{The CPEP contribution ratio as a function of the neutron-to-proton separation energy asymmetry ($\Delta S = S_{\rm p} - S_{\rm n}$) for the C, N, O, Si, Ar and Ca isotopic chains.}
	\label{e_sep}
  \vspace{-0.2cm}
\end{figure}

The interplay between $\sigma_{-i\text{n}}$ and $P_{i}^\text{evap.}$ results in an ``evaporation peak" at the isospin symmetric region, where the $\sigma_{\rm cc}^{\rm evap.}$ from CPEP plays the most significant role in contributing to the CCCS. Since the ``evaporation peak" is mainly shaped by the one-neutron removal as seen in Fig.~\ref{C_evap_xn_isospin} (b), another way to check the isospin dependence of the $\sigma _{\rm cc}^{\rm evap.}/\sigma _{\rm cc}^{\rm cor.}$ is to plot it as a function of the neutron-to-proton separation energy asymmetry ($\Delta S = S_{\rm p} - S_{\rm n}$, where $S_{\rm p}$ ($S_{\rm n}$) is the proton (neutron) separation energy). $\Delta S$ is a measure of the asymmetry of the Fermi surfaces in each nucleus. As shown in Fig.~\ref{e_sep}, the isotopic chains investigated here can already cover very wide neutron-proton separation energies and a similar ``evaporation peak" pattern is clearly seen in terms of $\Delta S$. The $\sigma _{\rm cc}^{\rm evap.}/\sigma _{\rm cc}^{\rm cor.}$ of Si, Ar, and Ca isotopes show the maxima at $\Delta S$ close to zero. The $\sigma _{\rm cc}^{\rm evap.}/\sigma _{\rm cc}^{\rm cor.}$ maxima of C, N, and O isotopes are a bit off from $\Delta S = 0$. This is because of the drastic change of their $S_{\rm n}$ with the isospin. It is interesting to investigate and discuss further features shown in Fig.~\ref{e_sep}, e.g., how the peak pattern will evolve up to heavier elements and if it is nuclear shell closure related or not.

For light nuclei up to Ca isotopes with $T_z \sim 0$, the protons and neutrons near the Fermi surface occupy identical orbitals, which allows for pairs consisting of a neutron and a proton, thus resulting in the enhancement of the binding energy, i.e., the ``Wigner Energy"~\cite{Wigner37}. Removing neutron(s) first from these nuclei will break the neutron-proton pair and release the enhanced binding energy, which leads to pre-fragments with relatively higher excitation energies and more likely results in the following charged-particle (most probably, proton) emission. Therefore, the nuclei at $T_z \sim 0$ are more sensitive to the CPEP. 
In addition, the ``Wigner Energy" decreases with the nuclear mass number and leads to the decrease in the $\sigma _{\rm cc}^{\rm evap.}/\sigma _{\rm cc}^{\rm cor.}$ ratio as seen for Si, Ar and Ca isotopes in Fig.~\ref{ratio_isospin} (a). Experimental data on nuclei with the extreme proton-neutron ratio, in particular on the proton-rich side, will be valuable to examine the peak-like pattern and further understand its evolution along the isotopic chain. 

We have also examined the reaction-energy dependence of the CPEP by performing calculations for the case of $^{40}$Ca +$^{12}$C. The $\sigma_{\rm cc}^{\rm evap.}$ values are estimated with EEDs from both the IQMD and the GS approaches. The $\sigma_{\rm cc}^{\rm evap.}$/$\sigma_{\rm cc}^{\rm cor.}$ magnitudes in both cases show a sharp increase of more than 20\% from 50 MeV/nucleon on and reach the peak values at around 200 MeV/nucleon, which accounts for 10\% of the CCCS. However, at higher energies, the results using EEDs from IQMD are more energy relevant compared with the GS case. 30\% difference in the ratios is observed from 700 MeV/nucleon on, which corresponds to 3\% of the CCCS. One needs to note, the IQMD model has consistently taken into account the reaction energy. Overall, unlike the systematic underestimation by the ZRGM, the $\sigma_{\rm cc}^{\rm cor.}$ predictions taking the CPEP contribution into account agree well with the available data~\cite{Yamaki13, Webber90}.

\section{Summary}
\label{sum}
In summary, we have measured thirteen CCCSs of $p$-shell isotopes on carbon at around 300 MeV/nucleon, in particular the new data on $^{11}$C, $^{13}$N and $^{15}$O extends our knowledge to the neutron-deficient side.
We concluded that the reaction mechanism beyond the Glauber model, like the CPEP following the projectile neutron removals, is essential to understand the existing experimental data at around 300 and 900 MeV/nucleon. We identified a universal ``evaporation peak" from C to Ca isotopic chains at the isospin symmetric region, where the CPEP plays a significant role in contributing to the CCCS. This peak pattern is reaction energy independent, but it is correlated to the nuclear structure in terms of the neutron-to-proton separation energy asymmetry. Such isospin-dependent CPEP feature also questions the method of deducing nuclear charge radii from the CCCSs, in which the isospin effect is not considered in calibrating the Glauber model. The reliable extraction of nuclear charge radii with the Glauber model requires the CPEP contribution to be subtracted from the experimental CCCS. This needs not only precise CCCS data but also requires a good estimation of the CPEP contribution. The latter relies on experimental measurements of the gamma de-excitation and the light particle emission from the pre-fragments for bench-marking the theory. Further measurements on nuclei with the extreme proton-neutron ratio up to the heavy mass region and of the evaporated particles near the reaction target can provide a more complete picture of the mechanism underlying and help to investigate questions like the equation of state of nuclear matter.

\section*{Acknowledgments}
We thank the HIRFL-CSR accelerator team for their efforts to provide a stable beam condition during the experiment. This work was supported partially by the National Key R\&D Program of China (Contract No. 2022YFA1602401), the National Natural Science Foundation of China (Nos. U1832211, 11922501, 11961141004, 12325506, 11475014, 11905260, and 12135004),  the Strategic Priority Research Program of Chinese Academy of Sciences (Grant No. XDB34010000), the Western Light Project of Chinese Academy of Sciences, and the Open Research Project of CAS Large Research Infrastructures.

\section{Supplemental Material}
The cross-section of only removing $i$-neutron is calculated as 
\begin{equation}
\begin{split}
\sigma_{-i\rm n}= 2\pi\binom{N_{\rm P}}{i} \int b\cdot T_{\rm p}(b)[t_{\rm n}(b)]^{N_{\rm p}-i}[1-t_{\rm n}(b)]^{i} db,
\end{split}
\label{eqin}
\end{equation}
where the binomial term is the number of combinations for removing $i$ neutrons out of the total $N_{\rm P}$ neutrons in the projectile and $b$ is the impact parameter. $T_{\rm p}(b)$ is the transparency function for projectile protons,
\begin{equation}
\begin{split}
T_{\rm p}(b)=\text{exp}[-\int ds\cdot\rho_{\rm p}^{\rm P}\cdot (\sigma_{\rm pp}\rho_{\rm p}^{\rm T}+\sigma_{\rm pn}\rho_{\rm n}^{\rm T})],
\end{split}
\label{eqin2}
\end{equation}
and $t_{\rm n}(b)$ is the probability that a single projectile neutron transmits the target density,
\begin{equation}
\begin{split}
t_{\rm n}(b)=\text{exp}[-\int ds\cdot(\rho_{\rm n}^{\rm P}/N_{\rm p})\cdot (\sigma_{\rm np}\rho_{\rm p}^{\rm T}+\sigma_{\rm nn}\rho_{\rm n}^{\rm T})].
\end{split}
\label{eqin3}
\end{equation}
Here, $s$ is the distance from the center of the projectile, $\rho_{\rm n(p)}^{\rm P(T)}$ represents the projectile (target) neutron (proton) densities integrated along the beam axis.
$\sigma_{ij}$ (with $i, j$ = p, n) is the nucleon-nucleon cross section.
The total neutron removal cross-section, $\sum_{i}\sigma_{-i\rm n}$, is normalized to the difference between the reaction cross-section and CCCS, both of which are calculated consistently with the Glauber model.

The corresponding charged-particle evaporation probability can be formulated as 
\begin{equation}
P_{i}^{\rm evap.}=\int_{S+V_b}^{\infty} \rho(E)\cdot w_e(E) dE \;.
\end{equation}
Here, $S$ is the charge-particle separation energy, $V_b$ is the particle Coulomb barrier, $\rho(E)$ is the excitation energy distribution (EED), and $w_e(E)$ is the probability to evaporate charged-particles at a certain excitation energy $E$. 

The Weisskopf-Ewing formalism~\cite{Weisskopf} is used to calculate the pre-fragment de-excitation via all cascade decays until there is insufficient excitation energy available for particle emission. $w_e(E)$ equals the probability of emitting at least one charged particle in the cascade decay process.

The EED of the pre-fragment is a key input to compute the CPEP. Gaimard-Schmidt (GS) approach is often used to estimate EEDs~\cite{GS91,Schei09}. There are two input parameters in the GS formula, the number of removed neutrons and the maximum excitation energy ($E_{\rm max}$) by removing one neutron. As a realistic solution, $E_{\rm max}$ could be adjusted to the experimental data. 

The $E_{\rm max}$ parameters of the GS type of EEDs used for the $\sigma _{\rm cc}^{\rm evap.}$ calculations are summarized in Table~\ref{tab2}. Adjusting these $E_{\rm max}$ parameters by up to 30\% can still reproduce the experimental cross sections within the uncertainties. For example, a $E_{\rm max}$ parameter of 36 MeV is used for the $\sigma _{\rm cc}^{\rm evap.}$ calculations of C isotopes to reproduce data measured at around 930 MeV/nucleon~\cite{Chulkov00} as shown in Fig.~\ref{evap_isospin_900}. In contrast, a $E_{\rm max}$ parameter of 16 MeV is used to reproduce the data of C isotopes measured at around 900 MeV/nucleon~\cite{Ritu16}. Such different $E_{\rm max}$ parameters employed in the $\sigma _{\rm cc}^{\rm evap.}$ calculations result from the systematic deviations between the two sets of experimental data. 


\begin{table}[htb]
\begin{threeparttable}[b]
\centering
\vspace{-0.2cm}
\caption{$E_{\rm max}$ parameters of the GS type of EEDs used for the $\sigma _{\rm cc}^{\rm evap.}$ calculations for C, N, O, Si, Ar and Ca isotopes shown in Fig.~\ref{f1_exp_data}, ~\ref{ratio_isospin} and ~\ref{evap_isospin_900}.}
\label{tab2}
\begin{tabular}{ p{1.8cm} p{2.5cm} p{2.5cm} }
\hline
\hline
$E_{\rm max}$ (MeV) & Isotopes & Energy (MeV/nucleon)\\
\hline
 35 \tnote{1} & C & 300   \\    
 34 \tnote{1} & N & 300   \\ 
 42 \tnote{1} & O & 300   \\
 39 \tnote{2} & Si & 300   \\    
 36 \tnote{2} & Ar & 300   \\ 
 55 \tnote{2} & Ca & 300   \\
 36 \tnote{3} & C & 930   \\    
 29 \tnote{3} & N & 930   \\ 
 13 \tnote{3} & O & 930   \\
 16 \tnote{3} & C & 900  \\    
 5  \tnote{3} & N & 900  \\ 
 5  \tnote{3} & O & 900  \\
\hline
\end{tabular}
  \end{threeparttable}
\begin{tablenotes}
       \item $^{1}$ for the calculations shown in Fig.~\ref{f1_exp_data} and Fig.~\ref{ratio_isospin}.
       \item $^{2}$ for the calculations shown in Fig.~\ref{ratio_isospin}.
       \item $^{3}$ for the calculations shown in Fig.~\ref{evap_isospin_900}.
     \end{tablenotes}

\end{table}

The isospin-dependent quantum molecular dynamics model (IQMD)~\cite{Su11} provides a microscopic-rooted EED, which takes into account the reaction dynamics and incident energy dependence. The IQMD coupled with the GEMINI model can well reproduce the reaction cross sections of $^{20}$Ne+$^{27}$Al, $^{56}$Fe+$^{27}$Al at hundreds of MeV~\cite{Su11} and the partial cross sections of $^{28}$Si+$^{12}$C at 218 MeV/nucleon~\cite{Li22}. Both the IQMD and GS models show a similar EED of the one-neutron removal pre-fragment for the same reaction system, but the IQMD model predicts broader EDDs of pre-fragments with more than one neutron removed.

\bibliographystyle{elsarticle-num}

\end{document}